\documentstyle[12pt,epsf]{article} 
\newcommand{\ba}{\begin{array}}
\newcommand{\ea}{\end{array}}
\newcommand{\bd}{\begin{displaymath}}
\newcommand{\ed}{\end{displaymath}}
\newcommand{\be}{\begin{equation}}
\newcommand{\ee}{\end{equation}}
\newcommand{\bea}{\begin{eqnarray}}
\newcommand{\eea}{\end{eqnarray}}

\def\g{\gamma}

\def\m{\mu}
\def\n{\nu}

\def\pr{\prime}

\begin{document}

\begin{flushright}
\tt{\bf arXiv:physics/0406024} 
\end{flushright}
\vskip 25pt
\begin{center}
{\Large {\bf Constancy of any signal velocity in all inertial frames}}
\vskip 20pt

\renewcommand{\thefootnote}{\fnsymbol{footnote}}

{\sf Rathin Adhikari $^{a}$
\footnote{E-mail address: adhikari2@satyam.net.in}}

\vskip 10pt  
 $^a$ {\it Jagadis Bose National Science Talent Search,
716, Jogendra Gardens, Kasba, Kolkata (Calcutta) 700078, India}

\vskip 15pt
{\bf Abstract}
\end{center}

\noindent
{\small 
If  isotropy of space and homogeneity
of space and time are the valid laws of nature then one can show that
velocity of any signal (whether it is light in a vacuum or in a
medium or something
having constant velocity in the rest frame of the observer and
by using which we can perform all (but not partial) measurements
of all physical quantities related with the particular experiment)
is constant
in all inertial frames. To verify this constancy for other signals apart
from light in vacuum (which is well-known) we have proposed
particularly one experiment  in which all measurements can be done using
light in water as signal.
If we  perform experiment in a medium then the signal connecting the
measurements of an event in two inertial frames can not be light in vacuum
rather it is light in the medium. In such cases where signal used
for measurement is not light in vacuum, relativistic relationships
of various physical quantities in Special Theory of Relativity (in which
light in vacuum has been used as signal) require
modifications.
This work might be verified from the 
analysis related with 
the number of cosmic muons reaching earth through atmosphere
or that related with the amount  of oscillation
of atmospheric neutrinos as  in both these cases signal for our measurement
can not
be light in vacuum but it is light in air.

\vskip 15pt
\begin{flushleft}
PACS NO. 03.30.+p     
\end{flushleft}

\newpage

\setcounter{equation}{0}
\setcounter{footnote}{0}
Einstein's Special theory of
relativity (STR) takes into account the important fact that signal velocity is
finite and the measurement of physical quantities in two inertial frames
depends upon this velocity. There are two   axioms in Einstein's 
STR.  One is that the laws of physics are same in all inertial frames  and
another  is the constancy of velocity of light (in vacuum) in all inertial
frames. Light in vacuum is the signal providing the link between the
measurements of physical quantities in inertial frames.  However, if we
perform experiment in a medium then the signal connecting the measurements
in two inertial frames can not be light in vacuum rather it is light
in the medium. Then two questions arise. Firstly how to modify the
relativistic relationships when signal is not light in vacuum and
secondly
whether only light in vacuum has a fundamental property that the velocity
of it  will
be constant in all inertial frames or this property is valid for other
signals also when  signal velocity is measured by using the signal itself
for the measurements
of length and time associated with the measurement of velocity
of the signal.

It may be noted that if one uses another signal instead of
light in vacuum (which is considered as signal in special theory of
relativity)  still clocks
in the rest frame of the observer measure  standard time. Synchronization
of two clocks in the observer's rest frame is same whatever signal we
use. This is because any signal one may use is supposed to have some
constant velocity
in the rest frame of the observer and one can appropriately adjust the
time of two clocks placed at two different places in the rest frame
by considering the time taken by that signal  to travel from one clock
to the other. In relating measurements of an event done in two
inertial frames light in
vacuum has been used in special theory of relativity. But
there are
situations when light in vacuum can not be the available signal. In   that
case it has been discussed in this letter, what should be the relationship
between the measurements in two inertial frames. 
We have discussed here that although clocks will measure
standard time in the rest frame of the observer but the relationship of
two measurements of the same event in two inertial frames will depend
upon the signal which has been used depending upon the  experiment.
Although measurements in the moving frame
using different signals are different but those can be easily related
provided that
we know the velocity
of different signals in the rest frame of the observer.
This has been discussed later.

We   have shown that any signal velocity remains constant
in different inertial frames where signal is that using which we
can perform all (not partial
\footnote{Suppose in part of the experiment light in vacuum is the
signal and in another part of the same experiment light in air is the
signal. In this case if one considers two inertial frames then
according to our work velocity  of the signal which is higher will
remain constant whereas the velocity of the signal which is lower
i.e., velocity of light in air will be found to change in different
inertial frames.})
measurements of all physical quantities corresponding
to a particular experiment. This has been derived by assuming that
isotropy of space and homogeneity of space and time are fundamental laws
of physics  for all inertial frames\footnote{It has also been assumed
that the relative velocity of two inertial frames is less than the
signal velocity which is discussed later.}.
General transformation of space
and time coordinates for two inertial frames for any signal which is
obtained is same as those in special theory
of relativity with only replacement of velocity of light in vacuum
in the transformation equations by the velocity of any signal considered
for all measurements of an
experiment.  At the end, to verify the constancy of velocity of the signal
(other than light in vacuum) we have proposed one experiment
in which light in
water has been considered as signal which is to be used for all
measurements in that experiment.
If velocity of light in water does
not change in another inertial frame (which we expect from our analysis)
then there will be no shift in the interference fringe in this experiment. 
Experiments  on cosmic muon decay  in the
atmosphere  or experiments on atmospheric neutrino oscillation
(where signal can not be light in
vacuum
rather it is light in air) might give some evidence
in support of the work presented here.

Without considering the constancy of velocity of light some authors
\cite{con} have obtained the relativistic transformations of space-time
coordinates in which
they have obtained a constant which on dimensionality ground have been
related with the velocity of light in vacuum which is known experimentally
as constant. In our approach in obtaining transformation rules for
space-time in different inertial frames we also have not consider the
constancy of velocity of light in vacuum. But we first prove that
not only light in vacuum but any signal velocity is constant
in all inertial frames. Then we obtain the transformation rules
which are valid  for any signal velocity.

An event  in one inertial frame  $S$ is characterized by the coordinates
$x, y, z, t$ and the same event as observed in another inertial frame
 $S^{\prime}$  is characterized by the coordinates $x^{\prime}, y^{\prime},
  z^{\prime}, t^{\prime}$. We consider that space and time are homogeneous
   and space is isotropic. Let
  $S^{\prime}$ is moving with velocity $v$ with respect to $S$ along the
  positive $x$ direction and
  we assume that the relative velocity
  $v$ of the two frames are along $x$ and $x^{\prime}$  axes.   Let us
  consider that the points $x=0$ and $x^{\prime}=0$  coincide when
  $t=0$ and also consider that $t^{\prime} =0  $  then.
  As space, time are considered as homogeneous so the
  transformation equations  are linear.  As  the $x$ axis coincides  always
  with the $x^{\prime}$ axis and  as the length of the same rod at rest
  in any inertial frames should be same, one can show that $y^{\prime}= y$
  and $z^{\prime}=z$. Now we have to find only $x^{\prime} $ and
  $t^{\prime}$
  in terms of the coordinates $x$ and $t$ of the event measured in the
  $S$ frame. Apart
  from considering $x=0$ we may
  consider $y =0$ and $z=0$ when $t=0$ without loosing any generality in our
  approach to find the transformation equations.

  Let us analyse  at some non-zero value  $t=T$ what is the value of $x$
  in the stationary frame
  corresponding to  moving frame point $x^{\prime}=0$. According to our
   consideration at
  $t=0$, point $x=0$ of $S$ frame coincides with $x^{\prime}=0$ of
  $S^{\prime}$ frame. Also according to our initial consideration
  the way $S^{\prime}$ is moving for that $x^{\prime} =0$ point
  will move with
  velocity $v$  with respect to $x=0$ point in the $S$ frame.
  So $x^{\prime}=0$ corresponds
  to $x=v \;T$.  Such correspondence is true for any value of $t$.
  Considering this and also taking into account that the
  transformation should be linear we can relate $x^{\prime}$ with any
  value of $x$ and $t$ in the stationary frame as
  \bea
  x^{\prime} = a\; \left( x - vt \right)
  \eea
  where $a$  is independent of $x$ and $t$.

  Next we like to analyse  at some non-zero
  value  $x=X$  what  is the value of $t$  in the stationary frame
  corresponding to moving frame time $t^{\prime}=0$.
  According to our consideration for an event occurring at $x=0$,
  the corresponding times in the $S$ frame and $S^{\prime}$ frame are
  $t=0$ and $t^{\prime}=0$ respectively.  To represent the distance
  $X$ from point $x=0$ where event has occurred  we imagine that
  two rods $AB$ and $A^{\prime}B^{\prime}$ are placed in the $S$ frame
  and $S^{\prime}$ frame respectively   as shown in Fig 1 (a).
  As observed in the $S$ frame at $t=0$ both are
  of equal length $X$ and  the end points of these rods $A$ and
  $A^{\prime}$ coincide with $x=0$ and $B$ and $B^{\prime}$ coincides
  with  $x=X$.  To find time at a distance $X$
  corresponding to the event occurred at $x=0$ and $t=0$
  at first we have to get common time for both these points or to
  synchronize the
  two clocks placed at these two points. For this
  we have to send signal
  from $x=0$ to $x=X$.   To synchronize the clocks
  we can set the time of the clock at $x=X$ to $X/v_s$ when the signal
  reaches $x=X$. Here $ v_s$ is the signal velocity in the rest frame
  \footnote{As stated earlier the synchronization of two clocks
  in the rest frame of the observer is independent of the
  signal which has been used.}.   According to our imagination
  it is like sending signal from one end $A$ to the other end $B$
    of the rod in the stationary frame
  to know $t=0$ and sending signal from one end $A^{\prime}$ to the other
   end $B^{\prime}$ of other rod placed in the moving frame
  to know $t $ value corresponding to $t^{\prime}=0$ of the moving frame.
   Signal starts at $t=0$ which is same as telling
   signal starts at $t^{\prime}=0$. When the  signal reaches $x=X$    i.e,
    $B$ of the other end of the rod the stationary frame observer finds
    that   his/her signal is yet to cover $X$ distance in the moving
    frame at that time i.e, yet to reach the other end  $B^{\prime}  $
    of the rod placed in the moving frame. The time taken by the
    signal to reach $x=X$ is $ X/v_s$. So when signal reaches $B$
    the distance between $B$ and $B^{\prime}$ is $v X/v_s$ which is
    shown in Fig 1 (b). The signal
    in the stationary frame will need extra time $v X/v_s^2$ to
    cover this distance\footnote{It is clear that for $v = v_s$, signal
     in the rest frame which starts at  $A^{\pr}$ at $t=0$
     will never reach $B^{\pr}$. In other words, we have assumed that
    $v < v_s$.  }. So although
    $t=0$ and $t^{\prime}=0$ are simultaneous but the signal
    reaching $X$ distance in two frames as observed by the stationary
    observer is not simultaneous and differs by $vX/v_s^2$ time. So
    at distance X, $t^{\prime} =0$ correspond to $t= v X/v_s^2$
    in the stationary frame.  Such correspondence is true for any
    other values
    of $x$ also. Considering this and and also taking into account
    that the transformation should be linear
   we can relate $t^{\prime}$ with any
  value of $t$ and $x$ in the stationary frame as
  \bea
  t^{\prime} = b\; \left( t - {vx \over v_s^2} \right)
  \eea
  where $b$  is independent of $x$ and $t$.

   Using eq. (1) \& (2) we can write $x$ in terms
   of $x^{\prime}$ and $t^{\prime}$ as
   \bea
   x= {1 \over  { a \left( 1- {v^2 \over v_s^2} \right)}} \left( x^{\prime}
   + {a \over b} v t^{\prime} \right)
   \eea
   If we consider the isotropy of space  and consider
   $S$ as the moving frame then it is 
   moving with $-v$ velocity    with respect to $S^{\prime}$ frame
   \cite{iso}.
   In that case 
   similar to equation (1) we can write
   \bea
   x= a^{\prime} \left( x^{\prime} + v t^{\prime} \right)
   \eea
  where $a^{\prime}$ is  independent of   $x^{\prime}$ and $t^{\prime}$.
  Comparing this with eq. (3)
  we find that
  \bea
  a=b\; ; \; \; \;
    a^{\prime} =
  {1 \over  { a \left( 1- {v^2 \over v_s^2} \right)}}
  \eea
   Using eq. (1) \& (2) we can write $t$ in terms
   of $t^{\prime}$ and $x^{\prime}$ as
   \bea
   t= {1 \over  { a \left( 1- {v^2 \over v_s^2} \right)}} \left( t^{\prime}
   + {b \over a} {v x^{\prime} \over v_s^2}\right)
   \eea
   Considering  $S$ as the moving frame,
   similar to equation (2) we can write
   \bea
   t= b^{\prime} \left( t^{\prime} + {v x^{\prime} \over v_s^2}\right)
   \eea
  where $b^{\prime}$ is  independent of   $x^{\prime}$ and $t^{\prime}$.
  Comparing this with eq. (6)
  we find that
  \bea
  a=b\; ; \;\;\; 
    b^{\prime} =
  {1 \over  { b \left( 1- {v^2 \over v_s^2} \right)}}
  \eea

  We  show here that the velocities of signal  as measured
  in both $S$
  and $S^{\prime}$  frame are same.      Let us consider that
  something is moving with velocity $u^{\prime}$  as measured
  in the $S^{\prime}$ frame and $u$ as measured in the $S$ frame.
  Then
  we can write
  \bea
  x^{\prime}=u^{\prime} t^{\prime}
  \eea
  and using eqs. (1) and (2)
  \bea
  x= {v +  {b  \over a} u^{\prime}  \over {1 + {b v \over a v_s^2}
  u^{\prime} } } \; t
  \eea
  But as $x = u t$ and $b=a$ so
  \bea
  u= {v +   u^{\prime}  \over {1 + 
 { u^{\prime} v \over v_s^2}}}  
  \eea
  which is the velocity addition rule. If $u^{\prime}$
  corresponds to signal velocity $v_s$ 
  then
  \bea
  u=v_s
  \eea
  So signal velocity  is
  same in $S$ and $S^{\prime}$ frames and  it can be shown that
  they are same
  in all inertial frames.
If $u^{\prime}=v_s$ then $u$ is also $v_s$ but not higher than that.
Suppose two signal
velocities $v_{s1}$ and $v_{s2}$ are available for our experiments.
If $v_{s1} < v_{s2}$ then one can see the change of $v_{s1}$ in two
inertial frames when one uses signal with velocity $v_{s2}$ although there
will  be no change in $v_{s1}$ when signal with velocity $v_{s1}$ is
used. So far apart from light in vacuum no other signal velocity has
been found to be constant in different inertial frames because in measuring
the velocity of that signal ($v_{s1}$ in this discussion) all related
physical quantities have not been
measured by the signal itself but by some other signal which is in general
light in vacuum ($v_{s2}$ in this discussion).   If in future
we find experimentally
any signal velocity  $v_{sh}$ which is higher than velocity of
light in vacuum in the observer's rest frame  then we shall find the
change in the velocity of light in vacuum in different inertial frames
provided that we do the
measurement of it by using the signal with $v_{sh}$.

  Here we  obtain the transformation rules for the space and time
  coordinates in two inertial frames corresponding to an event for any
  signal. For that we consider that to make things physically equivalent
   the length of the rod which is at rest in $S$ frame and measured in
   $S^{\prime}$ frame should be equal to the length of the same rod which
   is at rest in the $S^{\prime}$ frame and measured in the $S$ frame.
   The two ends of the rod is measured at the same time. Say the
   length of the rod is of one unit when measured in frame in which it
   is at rest. Then from equation (1) corresponding to the first measurement
   it follows that length is $a$ and from equation (4) corresponding
   to the second measurement  it follows that the length is $a^{\prime}$.
   Considering the isotropy of space these measurements should
   be physically equivalent and then
   \bea
   a= a^{\prime}.
   \eea
   Using
   this in eq.(5)
   we obtain
   \bea
   a={1 \over \sqrt{1 - {v^2 \over v_s^2}}}
   \eea
   Using eqs. (5), (8) and (13)
   \bea
   b^{\prime} =  a^{\prime} = a = b
   \eea
   Using this in eq. (8) we obtain
   \bea
   b={1 \over \sqrt{1 - {v^2 \over v_s^2}}}
   \eea
   So from eqs. (1), (2) (14) and (16) we obtain
   the general space time transformation rules  for any signal
   as
  \bea
   x^{\prime} =
   {1 \over \sqrt{1 - {v^2 \over v_s^2}}}
  \left( x - vt \right) \; ; \;\;\;
  t^{\prime} =
   {1 \over \sqrt{1 - {v^2 \over v_s^2}}}
    \left( t - {vx \over v_s^2} \right)
  \eea
These transformation rules  are like STR with only
replacement of the velocity of light in vacuum by the signal velocity
$v_s$.

We define the position-time four vector $x^{\mu} (v_s), \; \mu= 0,1,2,3$
corresponding to signal $v_s$ as $x^0 = v_s t, \; x^1 = x, \; x^2= y, \;
x^3 = z$ then the quantity which
remains invariant in different   inertial frame is
\bea
g_{\mu\nu}x^{\mu}(v_s)x^{\nu}(v_s)  =   g_{\mu\nu}x^{\mu\pr}(v_s)
x^{\nu\pr}(v_s).
\eea
Let us consider that there are two signals having constant velocity
 $v_{s1}$  and $v_{s2}$ respectively in the observer's rest frame. 
If the signal with velocity $v_{s2}$ is used for measurement in a particular
experiment then the Lorentz symmetry associated with $v_{s1}$ is broken
in that
experiment. Here
the Lorentz group is given by the transformations $x^2 = x^{2\pr},\;
x^3=x^{3\pr}$ and those given by eqs. (17)  and the invariance
in (18). Origin of the violation of Lorentz invariance in this work
is different from other works on this violation \cite{lv}. We differ
here in the sense that although Lorentz invariance corresponding to
one signal (say light in vacuum) is violated in the above-mentioned
experiment but the invariance corresponding to another signal will be
seen experimentally provided that all measurements are done using that
signal. Possible such experiments have been discussed later.

The measurements of physical quantities in $S$ and $S^{\prime}$ frames
as obtained by using different signals
can be  related. The measurement of relative velocity between two
different
inertial frames is independent of the signal which we use. Furthermore,
according to our earlier discussion this relative velocity is supposed
to be less than
all the velocities of different signals which might be used. If the
observed event is at rest with respect to say $S^{\prime}$ frame then
the measurement of $x^{\prime}$ and $t^{\prime}$  is independent
of the signal.  So if $x_{v_{s1}}$ and $t_{v_{s1}}$ are
the space and time coordinates
of an event occurred in $S^{\prime}$ frame and measured in $S$ frame
by using signal $v_{s1}$
and if $x_{v_{s2}}$ and $t_{v_{s2}}$  are the space and
time coordinates
of same event measured in the $S$ frame by using signal $v_{s2}$
then  from eqs. (17) it follows that 
\bea
x_{v_{s1}} = {\gamma_{v_{s1}} \over \gamma_{v_{s2}}} \; x_{v_{s2}}\; ;
\;\;  \;  \;
t_{v_{s1}} =v  \g_{v_{s1}}  \g_{v_{s2}}
\left[{ 1 \over  
v_{s1}^2  } -
{ 1 \over
v_{s2}^2  } \right]  x_{v_{s2}} +
{\g_{v_{s2}} \over \g_{v{s1}}  } \;t_{v_{s2}}
\nonumber
\eea
where $ \gamma_{v_{si}} = 1/\sqrt{1 - v^2/v_{si}^2}$.
If $v_{s2} > v_{s1}$ then $\gamma_{v_{s2}} < \gamma_{v_{s1}}$ and it follows
from above  equations that $x_{v{s1}} > x_{v_{s2}} $ for $x \neq 0$
and $t_{v{s2}} > t_{v{s1}} $ when $x_{v{s2}}=0$. 

Maxwell's equations for electromagnetic fields remain invariant
under the space time transformations   for the signal -
light in vacuum. One can show that these equations in a medium
remain invariant in which case light in the medium is the signal.
It is because in this case the
force transformation rules
between
two inertial
frames  are similar with the force transformation rules
of STR with only replacement of
velocity of light in
vacuum by the velocity of light in the medium in the transformation
rules. Hence the same replacement occurs  in the transformation
rules for electric and magnetic fields. Then there are  space time coordinates
in the Maxwell's equations in the transformation rules of which also
same replacement occurs.
Considering these transformations of 
space , time and electric and magnetic fields it can be
shown in a similar way like STR that Maxwell's equations in
a medium remain
invariant in other inertial frames when light in that medium
is used as signal. 

We discuss here how one may experimentally verify   whether  the
velocity of other signal apart from light in vacuum - say light
in water is constant in different inertial frames. The experimental set-up
has been shown diagrammatically in Figure 2.  There are four
mirrors $M_1, M_2, M_3$ and $M_4$ among which $M_1$ is partially
silvered and all are kept inside water.Monochromatic light coming
from the light source
falls on mirror $M_1$ and is then split into reflected and
transmitted parts and their directions are shown by arrows. The transmitted
and reflected light
follows some path where water is flowing and some path where water
is not flowing with respect to the apparatus as shown in the figure.
We may think that
the water which is flowing with respect to apparatus as the moving
frame $S^{\prime}$  and the water which is not flowing with respect
to apparatus and the apparatus as rest frame $S$. Waters in different
regions of the apparatus are of same refractive index. Part of light
reflected at mirror $M_1$ again reflects at mirror $M_4$, then mirror
$M_3$ and then mirror $M_2$ and finally reflects at $M_1$ again and go
to telescope. The telescope and the observer/detector are both inside water.
Another part of light after being transmitted at $M_1$ is reflected
at $M_2$,then $M_3$ after that $M_4$ and finally is transmitted
through  $M_1$ and go to telescope. So there will be interference due
to optical path difference of the transmitted and reflected beams and
fringe will be observed. If there is change in the velocity of light
in water which is flowing with respect to the velocity of light in
water which is not flowing with respect to the apparatus
then shift of fringe will be observed
with the variation of the  velocity of water flowing. However,
according to our work as only the light in water of particular
  refractive index
is used in the apparatus
as the signal there should not be any change in the velocity
of light in water which is flowing and so no shift in fringe should be
observed. If from the apparatus we remove all water which is not flowing
then the experiment
corresponds to the famous experiment of Fizeau in which fringe shift
was observed and is given by the formula
$\Delta N  \approx  4 l n^2 v_w \left( n^2 -1 \right)/(\lambda c n^2) $
where $2 l$ is the length of the path in which water is flowing, $n$
is the refractive index of water, $v_w$ is the velocity of water and
$c$ is velocity of light in vacuum and $\lambda$ is the wavelength of
light. Similar to light in a medium if one  use sound as signal for
all measurements in a particular experiment then the velocity
of sound would be same in all inertial frames.

Our work shows that the property that
the velocity of light in vacuum remains constant in all inertial frames
is not the property of it only rather in a general way one can tell
that any signal velocity is constant in different inertial frames.
When we use different signal other than light in vacuum then
in all transformation rules of special theory of relativity, velocity
of light $c$ in vacuum  is replaced by $v_s$. So it shows that
depending on the signal used whose velocity is $v_s$ the energy ,
momentum relation and mass, energy relation are given respectively by
\bea
E= v_s \sqrt{p^2 + m_0^2 v_s^2}\; ; \;\; \;  \;\;
E= m  v_s^2
\nonumber
\eea
where $m_0 $ is the rest mass of the particle.
So if we perform experiment of mass , energy relation
and if  whole experimental
set-up is fully covered in a medium where the velocity
of light is less than that in vacuum  we shall find lesser amount
of energy is released from a certain amount of mass in comparison
to the energy released from the same amount of mass in an
experiment in vacuum  or in an experiment whose set up is
fully covered with a medium where velocity
of light is higher than the earlier medium. This might be verified if
 experiment on $e^+e^-$ annihilation in a medium  could be performed
 and the energy of $\gamma$-rays is measured.

Half life for muon (moving in air) measured in air should be longer
than that predicted by STR  as half life of muon moving at velocity $v$
will be $ T(v) = {\left(1- v^2/v_s^2\right)}^{-1/2} \;\;T_0$
where $T_0$ is half life of muon at rest. Measuring the attenuation
of the cosmic ray muon beam as it proceeds down the atmosphere one
may verify this at different elevations. If one considers the
velocity of muon to be about 0.995 $c$ and
$v_s$  - the velocity of light in air about  $c/1.0002$ then half
life $T(v)$ of muon
in the atmosphere should be found experimentally to be  about
10.22 $T_0$. But this value according to STR is about 10.01 $T_0$.
Also for the atmospheric neutrinos one should
consider
light in air (instead of light in vacuum) as the signal in the
 relativistic relationships. Then there will be modifications in
 the decay width of pions and kaons 
 which decay to
 $\m^{\pm} + \n_{\mu} ({\bar \n_{\mu}})$
 and that of muon which decays to
 $e^{\pm} + \n_{e} ({\bar \n_{e}})$ as pion, kaon \& muon
 pass through the atmosphere. Analysing the different
 neutrino flux on the basis of this might give some evidence
 supporting  this work. 

If various experiments proposed in this work show negative results then
that                              
would indicate nature does not support the constancy of any signal velocity
in all inertial frames. In that case in the relativistic transformations
$c$ is
entering as some fundamental constant but not as signal velocity. Then
the concept of measurement in Special Theory of relativity
by using signal  is required to be changed.

\vskip 15pt

\noindent
{\bf Acknowledgment}

Author would like to thank Palash B. Pal for  comments and for
reading the manuscript and  theory division of
Saha Institute of Nuclear Physics
for kind hospitality during visit. Discussion with G. Rajasekaran has helped
in making further clarifications.
Author would like to thank Shilpi Ghosh for drawing the schematic
diagram and
Papiya Nandy for
granting academic leave.

\newpage
\begin{figure}[htb]
\mbox{}
\vskip 8.0in\relax\noindent\hskip 0.0in\relax
\includegraphics{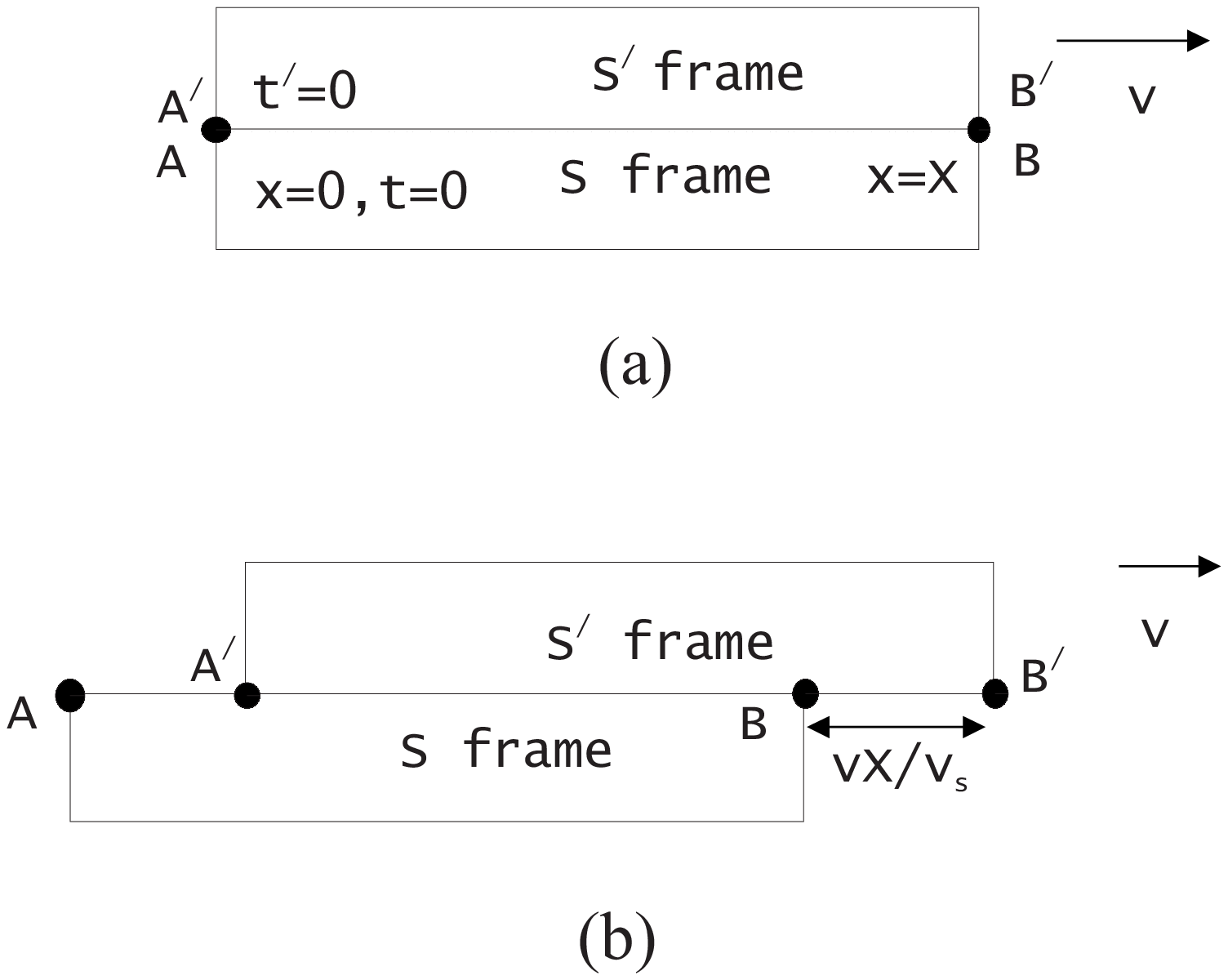} \vskip -2.25in
\caption{     In (a) two rods $AB$  and  $A^{\prime}B^{\prime}$
are shown
as observed in the $S$ frame when signal is sent from $x=0$
at $t=0$ when $t^{\prime}=0$ also. In (b)
two rods $AB$  and  $A^{\prime}B^{\prime}$ are shown
as observed in the $S$ frame when signal has reached $B$.  }

\end{figure}

\newpage
\begin{figure}[htb]
\mbox{}
\vskip 8.0in\relax\noindent\hskip -1.5in\relax
\includegraphics{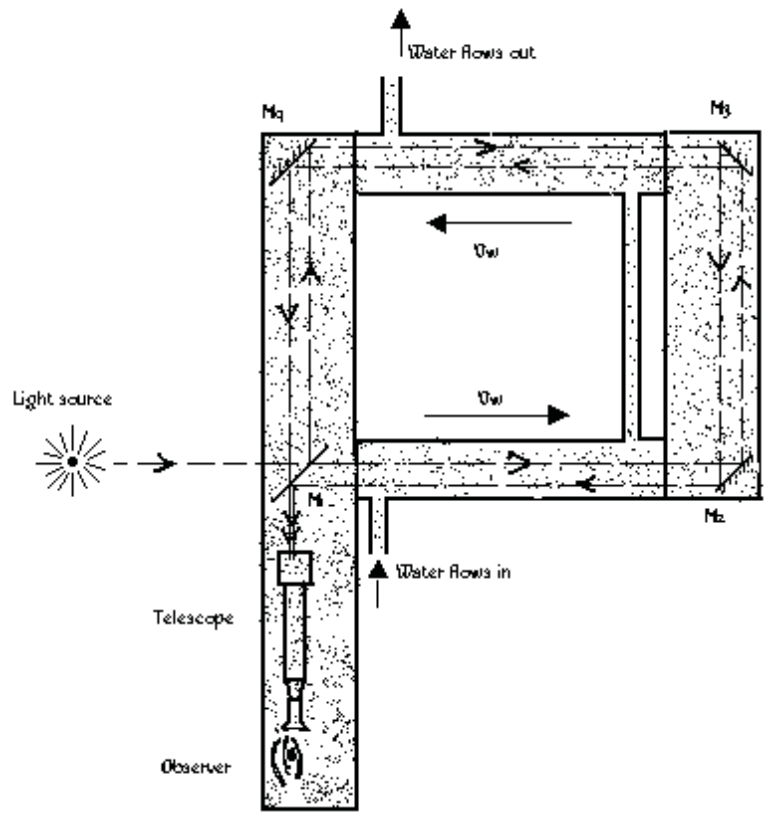} \vskip -3.25in
\caption{   Schematic experimental set-up to verify  the constancy
of the velocity of signal (light inside water)  .}
\end{figure}

\end{document}